**Original Paper**

# Topological photonic integrated circuits based on valley kink states

*Jingwen Ma, Xiang Xi, and Xiankai Sun*[*]

*Corresponding Author. E-mail: xksun@cuhk.edu.hk

Department of Electronic Engineering, The Chinese University of Hong Kong, Shatin, New Territories, Hong Kong

Valley pseudospin, a new degree of freedom in photonic lattices, provides an intriguing way to manipulate photons and enhance the robustness of optical networks. Here we experimentally demonstrated topological waveguiding, refracting, resonating, and routing of valley-polarized photons in integrated circuits. Specifically, we show that at the domain wall between photonic crystals of different topological valley phases, there exists a topologically protected valley kink state that is backscattering-free at sharp bends and terminals. We further harnessed these valley kink states for constructing high-$Q$ topological photonic crystal cavities with tortuously shaped cavity geometries. We also demonstrated a novel optical routing scheme at an intersection of multiple valley kink states, where light splits counterintuitively due to the valley pseudospin of photons. These results will not only lead to robust optical communication and signal processing, but also open the door for fundamental research of topological photonics in areas such as lasing, quantum photon-pair generation, and optomechanics.

# 1. Introduction

The field of photonic integrated circuits is gaining significant momentum because it allows cost-effective fabrication of nanophotonic devices and their seamless integration with microelectronics on a chip[1]. To date, achieving zero back reflection in arbitrarily shaped photonic circuits remains an outstanding challenge, which limits further increase of integration density of photonic networks[2]. Conventional wisdom of backscattering suppression relies on nonreciprocal devices[3], which are difficult to implement on integrated platforms due to the weak magneto-optical effect. Recently, inspired by the discovery of topological insulators in condensed-matter physics[4], photonic topological edge states[5-10] are considered as a promising strategy to suppress undesired backscattering. Despite early studies in nonreciprocal systems[11-15], photonic topological edge states without breaking time-reversal symmetry (TRS)[16-20] can be realized in systems that allow an analog of quantum spin-Hall effect[21, 22] with a pseudospin degree of freedom of photons. Various optical degrees of freedom have been used as pseudospin, such as polarizations in bianisotropic metamaterials[16], TE and TM modes in photonic crystals[18, 23, 24], chiralities of whispering-gallery modes in ring resonators[19, 25, 26], and more recently $p$- and $d$-orbitals in artificial photonic atoms[20, 27]. However, these schemes are either incompatible with integrated photonic platforms[16, 18, 23, 24] or suffer from bulky footprints[19] and large out-of-plane radiation loss[20, 27].

As a new binary degree of freedom labeled by different corners of the hexagonal Brillouin zone of two-dimensional Dirac materials[28-30], valley pseudospin provides an additional strategy to implement topologically robust transport in electronics[31, 32], acoustics[33], and photonics[34-36]. Instead of breaking TRS, reduction of spatial-inversion symmetry (SIS) can generate a nonvanishing valley-dependent Berry curvature and lead to quantum valley-Hall effect[28], which predicts the existence of a valley kink state at the domain wall between regions of different topological valley phases[37, 38]. Based on this valley kink state, various optical phenomena such as backscattering-immune refraction[34] and transport[35, 36], one-way Klein tunneling effect[39], and valley–spin locking[40] have been investigated theoretically[37-43] and experimentally[34, 44]. These valley-dependent phenomena, although indicating a promising strategy to suppress reflection without nonreciprocal media, have mostly been conducted in the radio-frequency regime with metallic materials[34, 35, 45]. Very recently, topological waveguiding of valley-polarized photons has been demonstrated on an integrated platform[46-48]. Harnessing the valley degree of freedom on a CMOS-compatible platform will not only revolutionize traditional integrated photonic devices for more robust and more compact optical communication networks, but also enable exploration of fundamental phenomena such as lasing[15, 25, 26], quantum photon-pair generation[49], and photon–phonon interactions[50, 51] in topologically nontrivial nanophotonic structures.

In this work, we exploit the valley degree of freedom to experimentally demonstrate topological waveguiding, refracting, resonating, and routing of photons on an integrated silicon photonic platform. More specifically, we realized a tightly confined valley kink state at the domain wall between two photonic crystals of different valley topological phases. By measuring the optical transmission along a tortuous topological domain wall, we validated that the valley kink state is backscattering-immune against sharp bends. We show that this valley kink state also exhibits topological protection when it is refracted into the surrounding photonic slab waveguides, facilitating integration with other optoelectronic devices on a chip. With the demonstrated valley kink states for waveguiding, we constructed geometry-independent optical cavities with topological protection, which support well-defined whispering-gallery modes with loaded $Q$ factors over $1.6 \times 10^4$. We also demonstrated a novel scheme of optical routing at an intersection of several topological domain walls, where the valley kink state splits counterintuitively as governed by the valley pseudospin degree of freedom. Our results represent a critical step to topologically protected integrated photonic circuits and devices by harnessing valley pseudospin of photons on a chip.

## 2. Results and Discussion

### 2.1. Topological bandgap with broken spatial inversion symmetry

We designed and fabricated photonic crystal structures on standard silicon-on-insulator wafers. As shown in Figure 1a, the photonic crystals are defined in a 220-nm silicon device layer on 3-μm-thick buried oxide. The photonic crystal adopts a honeycomb lattice with lattice constant $a = 520$ nm. Each unit cell includes two triangular air holes A and B with side length of $d_0 + \delta$ and $d_0 - \delta$, respectively (see Figure 1b). When these two triangular holes have the same size, i.e., $\delta = 0$, the photonic crystal has a $C_{6v}$ point symmetry. Setting different sizes of the triangles A and B with a nonvanishing $\delta$ breaks the spatial inversion symmetry (SIS) and reduces the point symmetry of the photonic crystal from $C_{6v}$ to $C_{3v}$. Here, the device pattern contains two different photonic crystal structures with opposite signs of $\delta$, which are labeled by the blue and red regions in Figure 1a. We will demonstrate that there naturally exists a valley kink state at the interface between these two regions, which will be used for constructing various topological integrated photonic devices.

Before further discussion about the valley kink states at the interface, we first study the bulk states of the photonic crystals. For the honeycomb photonic crystal, the valley degree of freedom is associated with two different high-symmetry points (K and K′) in the hexagonal first Brillouin zone (see Figure 1c). Here we only need to focus on the K valley because the TRS promises that the energy band diagram of the photonic crystal is always mirror-symmetric in the momentum space. Similar to graphene, the photonic crystals with equal size of triangular air holes A and B have a $C_{6v}$ point symmetry and support a gapless Dirac-cone band diagram. If the SIS is broken by a nonzero $\delta$, an energy bandgap will be opened at the K point, as numerically illustrated in Figure 1d. With tight-binding approximation, the Hamiltonian for the photonic states near the K and K′ points can be expressed as $\widehat{H}_{\tau_z} = v_D(\tau_z \sigma_x p_x + \sigma_y p_y) + \frac{\Delta}{2}\sigma_z$, where $\tau_z = 1\ (-1)$ indicates the K (K′) valley pseudospin, $v_D$ is related to the group velocity, $\mathbf{p} = (p_x, p_y)$ is the momentum vector relative to the K or K′ point, and $\sigma_{x,y,z}$ is the Pauli matrix (see Section S1, Supporting Information). The energy bandgap induced by broken SIS can be attributed to the spatial field distribution of the two eigenstates at the K point. As shown in Figure 1f, the $H_z$ component of one eigenstate at the K point is located in the upper triangular hole A, while the other eigenstate is located in the lower triangular hole B. Consequently, the size difference of the two triangles A and B lifts the degeneracy of these two eigenstates and induces a nonzero energy difference $\Delta$ between the two eigenfrequencies at the K point (see Figure 1e). It should be noted that the bandgap refers to that for the TE-like states near the K point, which are located in the bulk continuum of the TM-like modes. However, the slight TE–TM mode interaction caused by the buried oxide layer does not induce significant degradation in our devices' performance (see Section S3, Supporting Information).

The energy bandgap induced by broken SIS is topologically nontrivial. This can be verified by the nonvanishing Berry curvature defined in the momentum space. We numerically calculated the Berry curvature $\Omega(\mathbf{p})$ for the two bands with the parameters $d_0 = 347$ nm and $\delta = 28$ nm. As illustrated in Figure 1d, $\Omega(\mathbf{p})$ behaves like a source (sink) for the photonic states in the upper (lower) band near the K valley. For those photonic states in the opposite K′ valley, the distribution of $\Omega(\mathbf{p})$ takes an opposite value due to the TRS. This simulated distribution of $\Omega(\mathbf{p})$ also agrees well with our theoretical prediction (see Section S1, Supporting Information). Based on the Berry curvature, one can define a valley-indexed Chern number by integrating $\Omega(\mathbf{k})$ in half of the Brillouin zone surrounding the K or K′ valley, which takes a value of $C_{\pm,\tau_z} = \mp \frac{\tau_z}{2} \text{sgn}(\Delta)$ for the higher and lower bands in Figure 1d, respectively. The dependence of the valley Chern number on the sign of $\Delta$ indicates that the photonic structures with opposite values of $\delta$ belong to different topological phases, even though their energy band diagrams are identical.

## 2.2. Topologically robust valley kink states

At the domain wall between two photonic structures with opposite values of $\delta$, there naturally exists a valley kink state due to the bulk–edge correspondence[52]. This can be heuristically understood in the following way: when the valley pseudospin is conserved, the valley Chern number of the photonic crystal cannot change unless the bandgap is closed somewhere. This implies that when two photonic crystals with different valley Chern numbers are placed in contact, the energy bandgap must be closed at the interface, which leads to a valley kink state. It should be noted that not all types of domain walls can support a gapless valley kink state. For example, the armchair-type boundary does not support a topological edge state because it induces large scattering between the K and K′ valleys[53].

Here we focus on the zigzag-type boundary, as shown in Figure 2a, that theoretically conserves the valley pseudospin[53]. The photonic crystals in the blue and red regions on both sides of the domain wall possess the same parameters ($a$ = 520 nm, $d_0$ = 347 nm, and $|\delta|$ = 28 nm) except that $\delta$ is positive (negative) in the red (blue) region. Figure 2b plots the numerically simulated photonic band diagram of the line-defect structure shown in Figure 2a. In the reduced one-dimensional momentum space, an edge state (red dotted line) exhibits an almost linear dispersion relation in the bandgap frequency region. Consequently, the propagation direction of photons in the K (K′) valley is always positive (negative), considering the definition of group velocity $v_g = 2\pi\, df/dk_x$ where $f$ is the eigenfrequency of the state. It should be noted that the valley kink state is above the light line of the buried oxide layer (black dashed lines in Figure 2b). However, the buried oxide layer does not induce significant degradation in the propagation loss of the valley kink state (see Section S3, Supporting Information). Figure 2b also includes the other valley kink state (blue dotted line), which exists at the domain wall of a structure that has reversed sign of $\delta$ in Figure 2a, i.e., $\delta$ is negative (positive) in the red (blue) region. A recent work showed that this type of valley kink states can also be used for topological photonic waveguides[46]. Here, we will focus only on the structure where $\delta$ is positive (negative) in the red (blue) region.

We will show that the valley kink state is topologically robust against sharp bends[35]. As indicated by the yellow line in Figure 2c, the interface between the blue and red regions takes a bent geometry with eight sharp bends. At each corner the momentum of the guided photons rotates by $2\pi/3$. Intervalley scattering is suppressed at the $2\pi/3$ bends[37], so light can propagate smoothly without suffering from undesired backscattering. As shown in the simulated intensity distribution in Figure 2f, photons at the wavelength of 1565.0 nm are tightly confined to the domain wall and flow smoothly through the sharp bends. This topological robustness of the valley kink state was further experimentally confirmed by the measured transmission spectrum of the device. In addition to the device with a bent domain wall shown in Figure 2c, a device with a straight domain wall was also fabricated and measured as a control (see Section S3, Supporting Information). As shown in Figure 2e, in the bandgap region with wavelength from 1520 to 1575 nm, both the devices with a straight and bent domain wall exhibit high transmission with similar spectral dependency, while outside the bandgap region, the transmission of the device with a bent domain wall is much lower.

The measured transmission spectra agree well with the simulated results. This confirms our theoretical prediction that the valley kink state is topologically robust against sharp bends in the bandgap frequency range (marked in gray) in Figure 2e. Theoretically, the transmission spectra of the valley kink state should exhibit a flat top in the bandgap region. Here, the inclination of the transmission spectra toward the longer wavelength is attributed to wavelength-dependent coupling between the in/output waveguide and the valley kink state (See Section S2, Supporting Information).

We will show that the valley kink state is also topologically protected when it is refracted into the surrounding medium through a valley-preserving boundary[34]. As shown in Figure 2d, the boundary of the photonic crystal is of a zigzag type, which theoretically introduces zero intervalley scattering and preserves the valley pseudospin. Because the propagation direction of photons in the valley kink state is locked to their valley pseudospin, they cannot be reflected at the boundary but only refracted into the surrounding medium. The topologically protected refraction of the valley kink state is shown in the simulated optical field distribution at the wavelength of 1565.0 nm (Figure 2g). The angle of refraction $\theta$ in the surrounding medium is determined by conservation of momentum along the photonic crystal boundary. Instead of refracting light directly into the silicon slab, we designed a subwavelength structure as a low-index buffer to suppress the undesired high-order diffraction into the surrounding silicon slab (see Section S2, Supporting Information). The period of the subwavelength structure should be carefully chosen to avoid additional momentum that breaks the conservation of valley pseudospin. Here we set the period of the subwavelength buffer region to be $a/2 = 260$ nm. The simulated backscattering at the boundary of the photonic crystal is negligible (see Section S2, Supporting Information). The out-coupled light in the buffer region was then collected by a 5.5-μm-wide waveguide. As shown in Figure 2e, the measured coupling efficiency for each terminal is at least 40% in the wavelength range of 1520–1575 nm, which can be further enhanced by reducing the out-of-plane scattering and the spatial mismatch between the light distribution inside the buffer region and the fundamental mode of the waveguide. The topologically protected refraction at waveguide terminals facilitates integration of the valley kink states with other optoelectronic devices on a single chip.

**2.3. Tortuously shaped photonic cavities based on valley kink states**

Based on the valley kink states, we further constructed and experimentally demonstrated topologically protected photonic cavities with tortuous shapes. Figure 3a shows the device structure where a cavity is constructed by making a closed loop of the domain wall between the red ($d_0$ = 340 nm and $\delta$ = 55 nm) and blue ($d_0$ = 340 nm and $\delta$ = −55 nm) regions of the photonic crystal. Instead of having a continuous dispersion line across the bandgap as shown in Figure 2b, such configurations host valley-polarized traveling-wave modes with discrete eigenfrequencies. Due to the topological robustness of the valley kink states, the supported whispering-gallery modes with opposite chiralities are always orthogonal to each other despite the presence of sharp bends. To stimulate and detect these resonant states, we used traditional line-defect photonic waveguides as the input and output ports of the cavity, as shown in Figure 3b. The distance from the ports to the cavity was optimized such that the resonant photons can be detected efficiently. In addition to the device shown in Figure 3a, devices with different cavity geometries were also fabricated and measured (see Section S3, Supporting Information).

Figure 3d shows the measured spectra of optical transmission at the output port, where the peaks inside the bandgap region correspond to the resonant whispering-gallery modes. The inset shows a resonant mode near 1568.5 nm with a loaded $Q$ factor of $1.6 \times 10^4$. The measured intrinsic loss rate of the cavity is $\Gamma_i/2\pi$ = 9.88 GHz, and the measured extrinsic loss rate due to the in/output coupling is $\Gamma_e/2\pi$ = 2.07 GHz (see Section S3, Supporting Information). The simulated modal intensity distribution shown in Figure 3c indicates that the resonant photons are tightly confined to the topological domain wall, despite a tortuous cavity geometry with many sharp bends. We estimated the group index of the valley kink state with the measured optical transmission spectra. As shown in Figure 3e, the measured group index (blue dots) can be fitted with a parabolic curve (red line). This indicates that the chromatic dispersion is zero at the vertex of the parabola, where the free spectral range is a constant. This feature may find promising applications of, for example, topological quantum photon-pair generation by spontaneous four-wave mixing[54]. The measured group index of the valley kink state differs slightly from the value calculated from its band diagram (black dashed line). This might be attributed to the additional group delay induced by the sharp bends. After subtracting the influence of the bends, the measured group index (green line) is close to the simulated value (see Section S3, Supporting Information).

The realized photonic cavities are fundamentally different from the traditional topologically trivial cavities on photonic crystal platforms. More specifically, traditional photonic crystal cavities such as nanobeam cavities[55] and L3 defect cavities[56] usually have a regular modal profile and rely on careful structural optimization to reduce the loss of the resonant modes. The valley kink states, however, are under topological protection and thus insensitive to their geometry, which offers great flexibility in designing and fabricating cavities on photonic crystal platforms. This feature enables applications in, for example, dense wavelength division multiplexing technologies with topological robustness.

**2.4. Topological routing of valley kink states**

We further demonstrated a novel scheme of optical routing based on the topologically protected valley kink states. Traditionally, at an intersection of several photonic waveguides, light propagating in one waveguide is usually scattered into all the others. Intuitively, a sharper bend between the input and output waveguides leads to less light routed into that output channel. In our device, however, the topologically guided photons can be routed in a highly counterintuitive way.

In a geometry shown in Figure 4a, there are four topological domain walls (labeled by the yellow lines) separating the photonic crystals of opposite $\delta$ values with $|\delta|$ = 28 nm and $d_0$ = 347 nm. All these domain walls supporting the valley kink states intersect at one point. Figure 4b shows the zoomed-in structure near this intersection point. Figure 4d shows the simulated light field distribution at the wavelength of 1565.7 nm. When photons are injected into the device through Channel 1, they are selectively routed into Channels 2 and 4 but not into Channel 3, even though the deflection angle from Channel 1 to Channel 3 is much smaller than that from Channel 1 to Channels 2 and 4. This indicates that the valley kink state chooses to be routed through sharper bends in the device shown in Figure 4a.

This seemingly counterintuitive behavior of photons originates from the propagation direction of a valley kink state which is intrinsically locked to its valley pseudospin. As marked in Figure 4b, the topologically guided photons in the input Channel 1 belong to the K′ valley. Because the output Channels 2 and 4 have the same valley pseudospin as the input Channel 1, light can be freely routed. The valley kink state in the output Channel 3, however, belongs to the K valley, so that light cannot be coupled efficiently into this channel due to conservation of the valley pseudospin, which is manifested in the simulated optical field distribution in the momentum space.

As shown in Figure 4e, the input light at Port 1 stimulates the states near the K′ valley of the photonic crystal, which is then scattered into Ports 2 and 4.

Experimentally, this topological routing of valley kink states was confirmed by measuring the spectra of optical transmission from Channel 1 to Channels 2–4. Figure 4c shows that the measured output transmission to Channels 2 and 4 is at least 10 dB higher than that to Channel 3 in the wavelength range of 1530–1585 nm (labeled by the gray region). The reason for observing a tiny portion of light in Channel 3 is that the structural symmetry is slightly broken by the intersection point, causing some photons scattered from the K′ valley into the K valley. Compared with the simulated bandgap from 1520 to 1575 nm, the slight wavelength shift is caused by fabrication imperfection. The experimental results agree well with the simulated transmission shown in Figure 4f, confirming that the routing of photons is regulated by the valley pseudospin degree of freedom.

## 3. Methods

The energy band diagram and Berry curvature of the bulk states were obtained from three-dimensional simulation in MPB[57]. The other numerical results including the edge-state band diagram, the transmission spectra, and the optical field distributions were obtained from three-dimensional finite-difference time-domain simulation in Lumerical[58]. To obtain the projected band diagram in Figure 2b, we chose the size of the simulation region to be $a$ (= 520 nm) in the $x$ direction, $20\sqrt{3}a$ in the $y$ direction, and 4 μm in the $z$ direction. The Bloch boundary was applied in the $x$ direction, the continuous boundary was applied in the $y$ direction, and the perfectly-matched-layer boundary was applied in the $z$ direction. We placed a broadband dipole source near the silicon slab to excite all possible modes supported by the structure. We also set a monitor at the same place to record the optical field in the time domain. By Fourier transforming the recorded signal, one could find some sharp peaks in the frequency domain. Locating the center frequencies of these sharp peaks and sweeping the Bloch wavenumber yield the band diagram in Figure 2b.

The photonic crystals were fabricated with the nearby photonic waveguides and integrated grating couplers on an integrated silicon photonic platform. The device fabrication started from a standard silicon-on-insulator wafer manufactured by Soitec, with a 220-nm silicon device layer on 3-μm-thick buried oxide. The device patterns were defined by high-resolution electron-beam lithography and then transferred to the silicon device layer by plasma dry etching.

To measure the transmission spectra of the fabricated devices, light from a wavelength-tunable laser (Yenista Tunics T100s) was coupled via a single-mode fiber to the on-chip photonic waveguides through a two-dimensional integrated grating coupler which typically achieves a coupling efficiency of 25% (or 6 dB insertion loss) with 3-dB bandwidth of ~80 nm. The output signal was measured by a high-sensitivity optical power sensor (HP 81532A). The transmission spectra of the devices were obtained by sweeping the wavelength of the laser and measuring the corresponding transmitted optical power. We also fabricated grating couplers connected directly by strip waveguides on the same chip, so that their coupling efficiencies could be calibrated independently and their insertion loss could be subtracted from the transmission spectra of the devices.

## 4. Conclusion

In conclusion, we have experimentally demonstrated the valley kink states on an integrated photonic platform. Based on these valley kink states, we further realized topological photonic circuits with the functionalities of topological waveguiding, refracting, resonating, and routing of photons. Our work extends the traditional photonic integrated circuits and devices into the topologically nontrivial regime with demonstration of topologically protected photonic waveguides, microcavities, and beam splitters. Beyond the practical applications in robust optical communication and signal processing, our results have also opened the door for various fundamental optical phenomena in topologically nontrivial nanophotonic structures. For example, the large Kerr nonlinearity of silicon and the subwavelength light field confinement may lead to generation of topologically protected quantum photon pairs based on spontaneous four-wave mixing. Further optimization of the propagation loss and the chromatic dispersion of the valley kink states may lead to the emergence of optical frequency comb or even solitons that are topologically nontrivial. On the other hand, by removing the underlying silicon dioxide, the suspended photonic crystal structure is an ideal platform for research on interactions between photons and phonons under topological protection. Finally, realizing topologically protected photonic crystal waveguides and cavities in III–V compound semiconductors will lead to optical amplifiers without backscattering and lasers of tortuous cavity shapes.

**Supporting Information**

Additional supporting information may be found in the online version of this article at the publisher's website.

**Acknowledgements**

This work was supported by Hong Kong Research Grants Council Early Career Scheme (24208915), General Research Fund (14208717, 14206318), and NSFC/RGC Joint Research Scheme (N_CUHK415/15). The authors acknowledge valuable comments and suggestions from Professor Chang-Ling Zou, University of Science and Technology of China.

Received: ((will be filled in by the editorial staff))
Revised: ((will be filled in by the editorial staff))
Published online: ((will be filled in by the editorial staff))

**Keywords:** topological photonics, photonic integration, nanophotonic devices, silicon photonics, photonic crystals

# References

[1] L. A. Coldren, S. W. Corzine, M. L. Mashanovitch. *Diode Lasers and Photonic Integrated Circuits* (John Wiley & Sons, 2012).
[2] D. Dai, J. Bauters, J. E. Bowers. *Light Sci. Appl.* **2012**, *1*, e1.
[3] L. Bi, J. Hu, P. Jiang, D. H. Kim, G. F. Dionne, L. C. Kimerling, C. A. Ross. *Nat. Photonics* **2011**, *5*, 758.
[4] M. Z. Hasan, C. L. Kane. *Rev. Mod. Phys.* **2010**, *82*, 3045.
[5] L. Lu, J. D. Joannopoulos, M. Soljačić. *Nat. Photonics* **2014**, *8*, 821.
[6] A. B. Khanikaev, G. Shvets. *Nat. Photonics* **2017**, *11*, 763.
[7] T. Ozawa, H. M. Price, A. Amo, N. Goldman, M. Hafezi, L. Lu, M. Rechtsman, D. Schuster, J. Simon, O. Zilberberg. *arXiv:1802.04173* **2018**.
[8] W.-J. Chen, Z. H. Hang, J.-W. Dong, X. Xiao, H.-Z. Wang, C. T. Chan. *Phys. Rev. Lett.* **2011**, *107*, 023901.
[9] M. C. Rechtsman, J. M. Zeuner, Y. Plotnik, Y. Lumer, D. Podolsky, F. Dreisow, S. Nolte, M. Segev, A. Szameit. *Nature* **2013**, *496*, 196.
[10] A. Slobozhanyuk, S. H. Mousavi, X. Ni, D. Smirnova, Y. S. Kivshar, A. B. Khanikaev. *Nat. Photonics* **2016**, *11*, 130.
[11] F. D. M. Haldane, S. Raghu. *Phys. Rev. Lett.* **2008**, *100*, 013904.
[12] Z. Wang, Y. D. Chong, J. D. Joannopoulos, M. Soljačić. *Phys. Rev. Lett.* **2008**, *100*, 013905.
[13] Z. Wang, Y. Chong, J. D. Joannopoulos, M. Soljačić. *Nature* **2009**, *461*, 772.
[14] K. Fang, Z. Yu, S. Fan. *Nat. Photonics* **2012**, *6*, 782.
[15] B. Bahari, A. Ndao, F. Vallini, A. El Amili, Y. Fainman, B. Kanté. *Science* **2017**, *358*, 636.
[16] A. B. Khanikaev, S. Hossein Mousavi, W.-K. Tse, M. Kargarian, A. H. MacDonald, G. Shvets. *Nat. Mater.* **2012**, *12*, 233.
[17] R. O. Umucalilar, I. Carusotto. *Phys. Rev. A* **2011**, *84*, 043804.
[18] T. Ma, A. B. Khanikaev, S. H. Mousavi, G. Shvets. *Phys. Rev. Lett.* **2015**, *114*, 127401.
[19] M. Hafezi, E. A. Demler, M. D. Lukin, J. M. Taylor. *Nat. Phys.* **2011**, *7*, 907.
[20] S. Barik, A. Karasahin, C. Flower, T. Cai, H. Miyake, W. DeGottardi, M. Hafezi, E. Waks. *Science* **2018**, *359*, 666.
[21] C. L. Kane, E. J. Mele. *Phys. Rev. Lett.* **2005**, *95*, 226801.
[22] B. A. Bernevig, T. L. Hughes, S.-C. Zhang. *Science* **2006**, *314*, 1757.
[23] A. Slobozhanyuk, A. V. Shchelokova, X. Ni, S. H. Mousavi, D. A. Smirnova, P. A. Belov, A. Alù, Y. S. Kivshar, A. B. Khanikaev. *arXiv:1705.07841* **2017**.
[24] K. Lai, T. Ma, X. Bo, S. Anlage, G. Shvets. *Sci. Rep.* **2016**, *6*, 28453.
[25] G. Harari, M. A. Bandres, Y. Lumer, M. C. Rechtsman, Y. D. Chong, M. Khajavikhan, D. N. Christodoulides, M. Segev. *Science* **2018**, *359*, eaar4003.
[26] M. A. Bandres, S. Wittek, G. Harari, M. Parto, J. Ren, M. Segev, D. N. Christodoulides, M. Khajavikhan. *Science* **2018**, *359*, eaar4005.
[27] L.-H. Wu, X. Hu. *Phys. Rev. Lett.* **2015**, *114*, 223901.
[28] D. Xiao, W. Yao, Q. Niu. *Phys. Rev. Lett.* **2007**, *99*, 236809.
[29] H. Zeng, J. Dai, W. Yao, D. Xiao, X. Cui. *Nat. Nanotechnol.* **2012**, *7*, 490.
[30] J. R. Schaibley, H. Yu, G. Clark, P. Rivera, J. S. Ross, K. L. Seyler, W. Yao, X. Xu. *Nat. Rev. Mater.* **2016**, *1*, 16055.

[31] L. Ju, Z. Shi, N. Nair, Y. Lv, C. Jin, J. Velasco Jr, C. Ojeda-Aristizabal, H. A. Bechtel, M. C. Martin, A. Zettl, J. Analytis, F. Wang. *Nature* **2015**, *520*, 650.
[32] L.-J. Yin, H. Jiang, J.-B. Qiao, L. He. *Nat. Commun.* **2016**, *7*, 11760.
[33] J. Lu, C. Qiu, L. Ye, X. Fan, M. Ke, F. Zhang, Z. Liu. *Nat. Phys.* **2016**, *13*, 369.
[34] F. Gao, H. Xue, Z. Yang, K. Lai, Y. Yu, X. Lin, Y. Chong, G. Shvets, B. Zhang. *Nat. Phys.* **2017**, *14*, 140.
[35] X. Wu, Y. Meng, J. Tian, Y. Huang, H. Xiang, D. Han, W. Wen. *Nat. Commun.* **2017**, *8*, 1304.
[36] J. Noh, S. Huang, K. P. Chen, M. C. Rechtsman. *Phys. Rev. Lett.* **2018**, *120*, 063902.
[37] M. Tzuhsuan, S. Gennady. *New J. Phys.* **2016**, *18*, 025012.
[38] T. Ma, G. Shvets. *Phys. Rev. B* **2017**, *95*, 165102.
[39] X. Ni, D. Purtseladze, D. A. Smirnova, A. Slobozhanyuk, A. Alù, A. B. Khanikaev. *Sci. Adv.* **2018**, *4*, eaap8802.
[40] J.-W. Dong, X.-D. Chen, H. Zhu, Y. Wang, X. Zhang. *Nat. Mater.* **2016**, *16*, 298.
[41] X.-D. Chen, F.-L. Zhao, M. Chen, J.-W. Dong. *Phys. Rev. B* **2017**, *96*, 020202.
[42] Y. Yang, H. Jiang, Z. H. Hang. *Sci. Rep.* **2018**, *8*, 1588.
[43] O. Bleu, D. D. Solnyshkov, G. Malpuech. *Phys. Rev. B* **2017**, *95*, 235431.
[44] H. Xue, F. Gao, Y. Yu, Y. D. Chong, G. Shvets, B. Zhang. *arXiv:1811.00393* **2017**.
[45] D. Song, V. Paltoglou, S. Liu, Y. Zhu, D. Gallardo, L. Tang, J. Xu, M. Ablowitz, N. K. Efremidis, Z. Chen. *Nat. Commun.* **2015**, *6*, 6272.
[46] M. I. Shalaev, W. Walasik, A. Tsukernik, Y. Xu, N. M. Litchinitser. *Nat. Nanotechnol.* **2018**, *14*, 31.
[47] X.-T. He, E.-T. Liang, J.-J. Yuan, H.-Y. Qiu, X.-D. Chen, F.-L. Zhao, J.-W. Dong. *Nat. Commun.* **2019**, *10*, 872.
[48] T. Yamaguchi, O. Yasutomo, R. Katsumi, K. Watanabe, S. Ishida, A. Osada, Y. Arakawa, S. Iwamoto. *Appl. Phys. Express* **2019**, *12*, 062005.
[49] S. Mittal, E. A. Goldschmidt, M. Hafezi. *Nature* **2018**, *561*, 502.
[50] V. Peano, C. Brendel, M. Schmidt, F. Marquardt. *Phys. Rev. X* **2015**, *5*, 031011.
[51] M. J. Ablowitz, C. W. Curtis, Y.-P. Ma. *Phys. Rev. A* **2014**, *90*, 023813.
[52] Y. Hatsugai. *Phys. Rev. B* **1993**, *48*, 11851.
[53] Y. Ren, Z. Qiao, Q. Niu. *Rep. Prog. Phys.* **2016**, *79*, 066501.
[54] J. W. Silverstone, D. Bonneau, K. Ohira, N. Suzuki, H. Yoshida, N. Iizuka, M. Ezaki, C. M. Natarajan, M. G. Tanner, R. H. Hadfield, V. Zwiller, G. D. Marshall, J. G. Rarity, J. L. O'Brien, M. G. Thompson. *Nat. Photonics* **2013**, *8*, 104.
[55] P. B. Deotare, M. W. McCutcheon, I. W. Frank, M. Khan, M. Lončar. *Appl. Phys. Lett.* **2009**, *94*, 121106.
[56] Y. Akahane, T. Asano, B.-S. Song, S. Noda. *Opt. Express* **2005**, *13*, 1202.
[57] S. G. Johnson, J. D. Joannopoulos. *Opt. Express* **2001**, *8*, 173.
[58] https://www.lumerical.com/

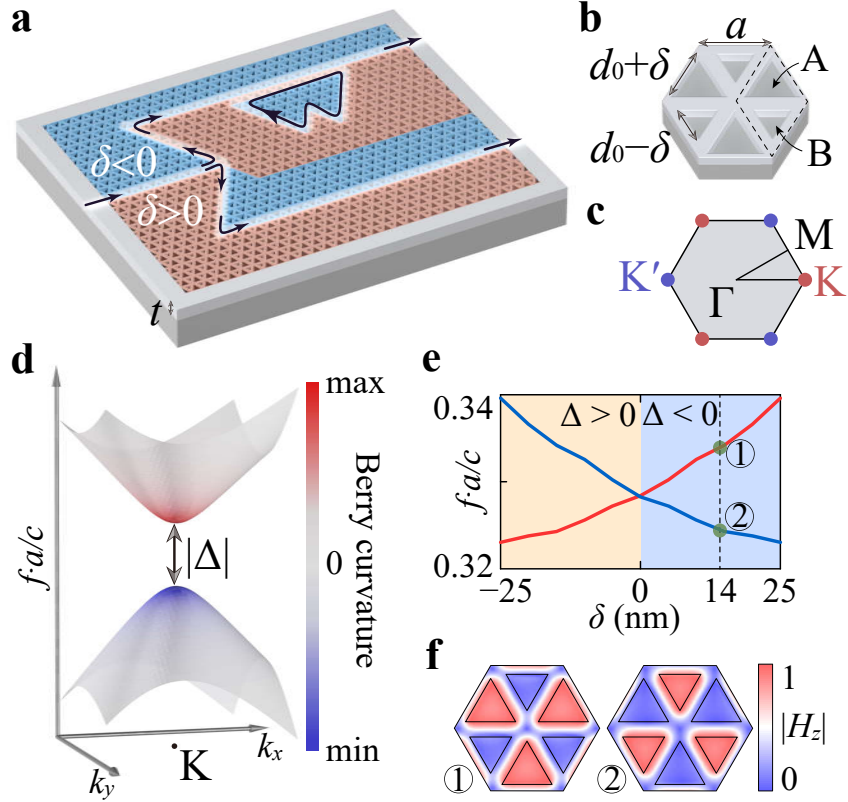

**Figure 1.** Topological bandgaps with broken spatial inversion symmetry. (a) Illustration of a topological photonic integrated circuit based on valley kink state. The valley kink state exists at the domain wall between two types of photonic crystals with $\delta$ of opposite signs. (b) Unit cell of the photonic crystal containing two types of triangular holes A and B, which have different sizes with a nonzero $\delta$. (c) The first Brillouin zone of the photonic crystal with two types of high-symmetry points K and K′ located at the corners. (d) Energy band diagram and the Berry curvature of the two lowest bands near the K valley, showing opposite signs of the Berry curvature for the two bands. (e) Eigenfrequencies of the two lowest bands at the K point with varying $\delta$. $\Delta$ denotes the difference between these two eigenfrequencies, and the sign of $\Delta$ is always opposite to that of $\delta$. (f) Modal field distributions ($H_z$ component) of these two states.

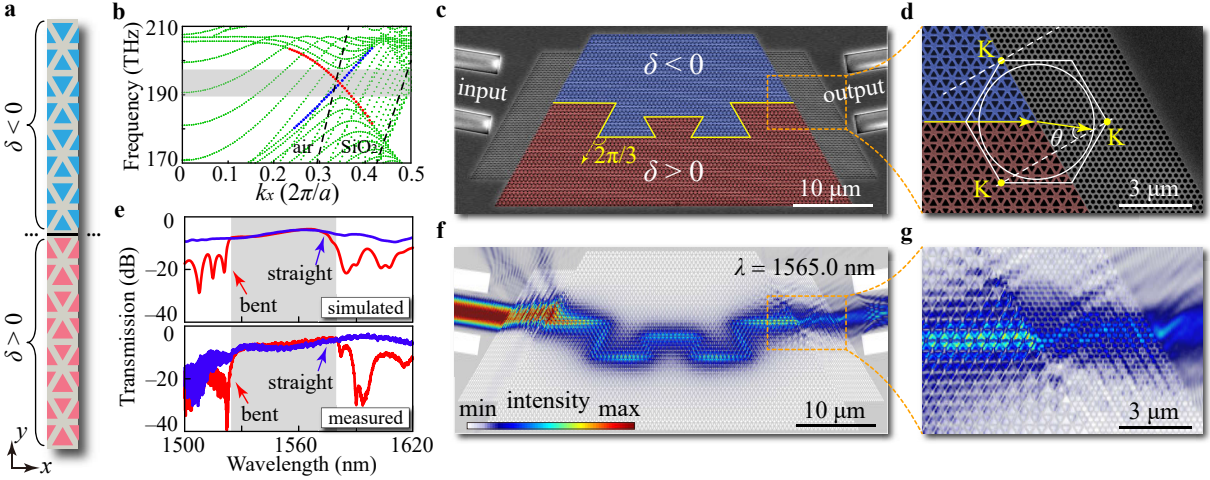

**Figure 2.** Topologically protected propagation and refraction of the valley kink state. (a) Structure of the topological domain wall that supports the valley kink state. (b) Projected band diagram of the valley kink state. The red dotted line represents the valley kink state at the domain wall shown in (a). The blue dotted line represents the other valley kink state for a structure that has reversed sign of $\delta$ in (a). The black dashed lines represent the light lines of the air and silicon dioxide claddings. (c) Scanning electron microscope (SEM) image of the device that supports backscattering-immune propagation of light against sharp bends. The photonic crystals in the blue and red regions take opposite values of $\delta$. The domain wall between them contains eight $2\pi/3$ bends. (d) SEM image showing a terminal of the valley kink state. The photons guided in the valley kink state exhibit topologically protected refraction, meaning that light can only be refracted into the buffer region and cannot be scattered back into the counterpropagating state. The white hexagon indicates the boundary of the first Brillouin zone of the photonic crystal, and the white circle delineates the momentum of light in the buffer region. The direction of refraction is determined by $\theta$. (e) Simulated and measured transmission spectra of devices with a straight and bent domain wall. In the bandgap region (marked in gray), both types of the devices exhibit high transmission with similar spectral dependency. (f), (g) Simulated intensity distribution at the wavelength of 1565.0 nm. Light is confined to and guided by the topological domain wall, exhibiting backscattering-immune propagation at sharp bends and topological refraction at terminals.

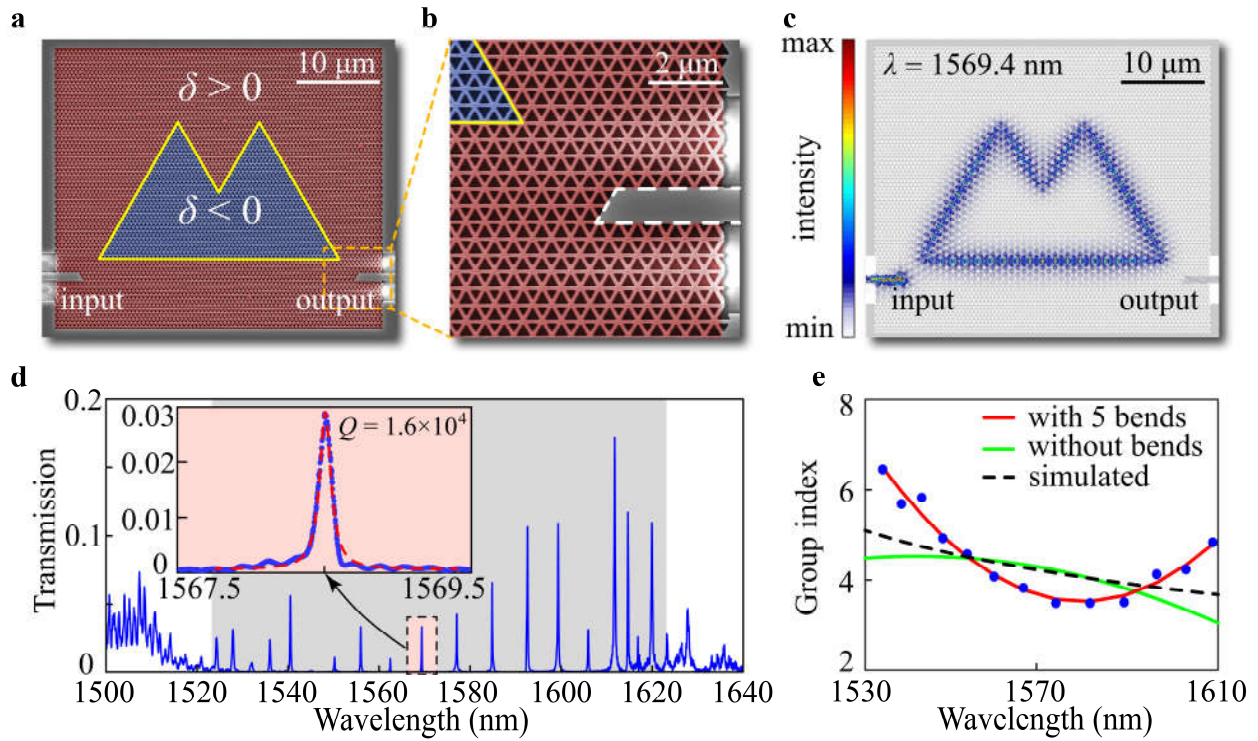

**Figure 3.** Tortuously shaped photonic cavity with topological robustness. (a) SEM image of the topological photonic cavity. The domain wall between the red ($\delta > 0$) and blue ($\delta < 0$) regions forms a closed loop. (b) SEM image showing the output port of the topological photonic cavity. (c) Simulated optical intensity distribution at the wavelength of 1569.4 nm. Light is stimulated at the input port and collected at the output port. (d) Measured optical transmission spectra of the device. Many whispering-gallery modes are located in the bandgap region (marked in gray). The zoomed-in spectrum shows a loaded $Q$ factor of $1.6 \times 10^4$ for a resonant mode at the wavelength of ~1568.5 nm. (e) Group-index spectra of the valley kink state. The measured data (blue dots) can be fitted with a parabolic curve (red line). The chromatic dispersion is zero at the vertex of this parabola. The green line plots the experimentally derived group index after subtracting the influence of the bends in the cavity (see Section S3, Supporting Information). The black dashed line plots the group index calculated from the band diagram of the valley kink state.

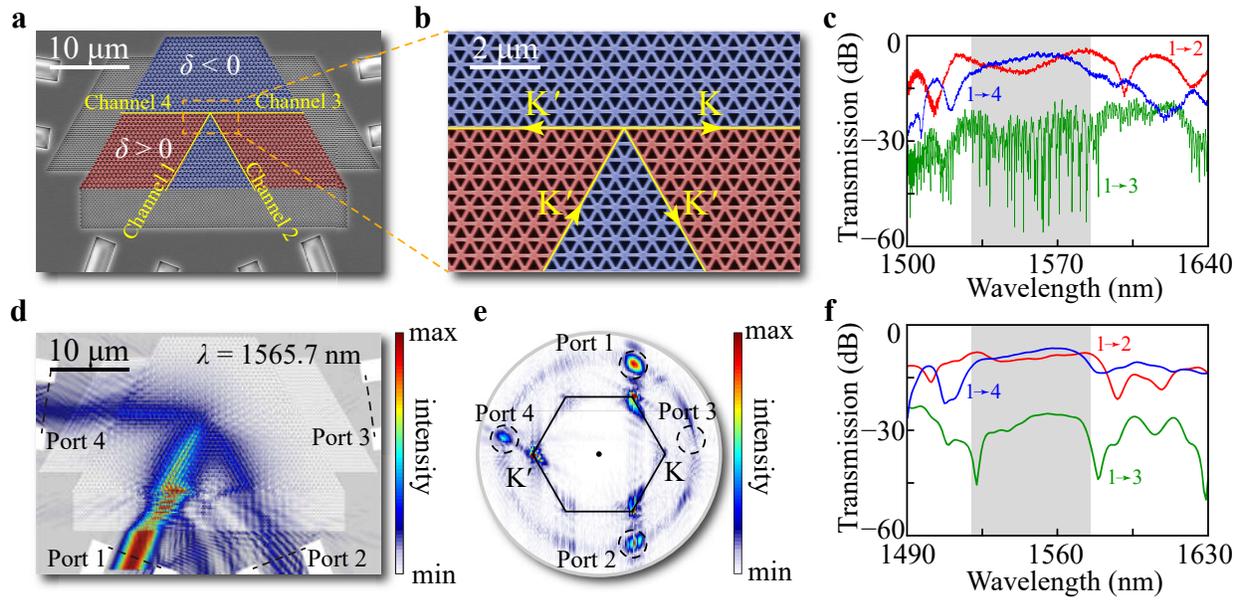

**Figure 4.** Topological routing of the valley kink state. (a) SEM image of the device that supports topologically protected routing of the valley kink state. (b) Zoomed-in image showing the intersection point of four topological channels. (c) Measured and (f) simulated spectra of optical transmission from Port 1 to Ports 2–4, which are plotted respectively as the red, green, and blue lines. (d) Simulated optical intensity distribution at the wavelength of 1565.7 nm. Light injected into Port 1 is routed into Ports 2 and 4, but not into Port 3. (e) Simulated optical intensity distribution in the momentum space. The power injected in Port 1 excites the photonic states near the K′ valley, which are then routed into Ports 2 and 4.

**Graphical Abstract:**

We have proposed and experimentally demonstrated a new type of topological edge states on an integrated photonic platform, which enables various functions of topological light manipulation, including waveguiding, refracting, resonating, and routing of photons. These phenomena correspond to a portfolio of backscattering-free integrated photonic components such as photonic waveguides, high-$Q$ microcavities, and optical beam splitters, all of which are essential for the development of a large-scale topological photonic system.

**ToC Figure:**

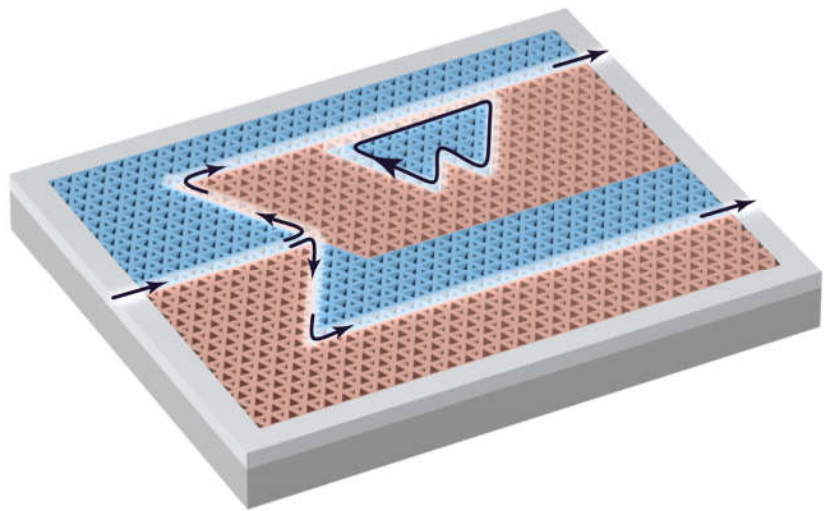